\documentclass{Rinton-P9x6}
\usepackage{amssymb}
\usepackage{amsmath}

\def\Cas{\mathrm{Cas}}
\def\pl{\mathrm{P}}
\def\T{\mathrm{T}}
\def\B{\mathrm{B}}
\def\D{\mathrm{D}}
\def\bk{\mathbf{k}}
\def\absk{\left|\mathbf{k}\right|}
\def\dd{\mathrm{d}}
\def\calP{\mathcal{P}}
\def\Lif{\mathrm{Lifshitz}}
\def\phit{\widetilde{\phi}}
\def\TE{\mathrm{TE}}
\def\TM{\mathrm{TM}}
\def\kappam{K}

\begin{document}

\title{The Casimir force between real mirrors \\
at non zero temperature}

\author{Serge Reynaud and Astrid Lambrecht}

\address{Laboratoire Kastler Brossel, UPMC case 74, Campus Jussieu, \\
F-75252 Paris cedex 05 France}  

\author{Cyriaque Genet}

\address{Huygens Laboratory, Universiteit Leiden, \\
P.O. Box 9504, 2300 RA Leiden The Netherlands}

\date{December 1st, 2003}

\maketitle

\abstracts{
The Casimir force between dissipative metallic mirrors at non zero temperature 
has recently given rise to contradictory claims which have raised doubts about
the theoretical expression of the force. In order to contribute to the resolution
of this difficulty, we come back to the derivation of the force from basic 
principles of the quantum theory of lossy optical cavities. 
We obtain an expression which is valid for arbitrary mirrors, including 
dissipative ones, characterized by frequency dependent reflection amplitudes. 
}

\section{Introduction}

The Casimir force  \cite{Casimir48} is the most accessible effect of vacuum 
fluctuations in the macroscopic world. Since vacuum energy raises difficulties 
at the interface of quantum and gravitational theories, 
it is worth testing this effect with the greatest care \cite{Reynaud01}. 
Recent measurements with a precision near the \% level 
\cite{Harris00,Chan01,Decca03prl,Bordag01,Lambrecht02} 
allow an accurate comparison between measured values and theoretical 
predictions, used for example in the search for new forces predicted in 
theoretical unification models \cite{Decca03prd}.
In this context, it is extremely important to account for the conditions
of real experiments.

Casimir calculated the force between perfectly reflecting mirrors in vacuum,
that is at zero temperature. This is why he found an expression for the force 
$F_\Cas$ which only depends on the distance $L$, the area $A \gg L^2$ and two 
fundamental constants, the speed of light $c$ and Planck constant $\hbar$, 
\begin{eqnarray}
F_\Cas &=& \frac{\hbar c\pi ^2 A}{240L^4}  \label{eqCasimir}
\end{eqnarray}
But the experiments performed with metallic mirrors are affected by the effect of
imperfect reflection at distances smaller than a few plasma wavelengths 
\cite{Lifshitz56,Schwinger78,Lambrecht00}.
For experiments at room temperature, the effect of thermal field fluctuations,
superimposed to that of vacuum, affects the Casimir force at distances larger than 
a few microns \cite{Mehra67,Brown69,Genet00}.
Most experiments are also performed between a plane and a sphere 
with the force estimation involving a geometry correction.
In the present paper, we consider the original Casimir geometry with 
perfectly plane and parallel mirrors and we restrict our attention on the 
correlated effects of imperfect reflection and non zero temperature. 

Recent publications have given rise to contradictory estimations of the Casimir 
force between dissipative mirrors at non zero temperature
\cite{Bostrom00,Svetovoy00,Bordag00,Klimchitskaya01,Hoye03}. 
This polemical discussion has raised doubts about the validity of the 
Lifshitz' formula often used to evaluate the Casimir force between imperfect mirrors. 
It has even led some authors to question the applicability of the scattering
description of mirrors in the presence of losses  \cite{Klimchitskaya01}. 
As shown below, this questioning certainly goes too far. 
In this short contribution, we come back to the basic 
derivation of the Casimir force between arbitrary mirrors at finite temperature. 
We extend to non zero temperature the derivation already given
for (zero temperature) vacuum fluctuations in \cite{Genet03}. 
Focusing our attention on general expressions written in terms of reflection 
amplitudes, we obtain formulas which were already known \cite{Jaekel91} for 
lossless mirrors but are shown here to be valid for lossy mirrors as well. 

\section{The Casimir force as an integral over real frequencies}

Assuming thermal equilibrium at temperature $T$, we now repeat the 
main steps of the derivation presented in \cite{Genet03} for $T=0$.
The main idea of this derivation is to evaluate the radiation pressure 
of fields on the two mirrors by using scattering theory \cite{Jaekel91}. 

The field modes are conveniently characterized by quantum numbers preserved 
throughout the scattering processes, namely the frequency $\omega$, 
the transverse wavevector $\bk$ with components $k_{x},k_{y}$ in the 
plane of the mirrors and the polarization $p$.
On each of the two mirrors, the scattering couples modes with the 
same values for $\omega$, $\bk$ and $p$ but opposite values
for the longitudinal wavevector $k_z$.
We denote by $\left(r_\bk^p\left[\omega\right]\right)_j$ the reflection amplitude of
the mirror $j=1,2$ as seen from the inner side of the cavity. 
This scattering amplitude obeys 
general properties of causality, unitarity and high frequency transparency.
The additional fluctuations accompanying losses inside the mirrors are
deduced from the optical theorem applied to the scattering process which 
couples the modes of interest and the noise modes \cite{Barnett96,Courty00}.

The loop functions which characterize the optical response of the cavity 
to an input field play an important role in the following
\begin{eqnarray}
&&f_\bk^{p}\left[\omega\right] = 
\frac{\rho_\bk^p\left[\omega\right]}{1-\rho_\bk^p\left[\omega\right]}
\quad,\quad \rho_\bk^p\left[\omega\right] = 
\left(r_\bk^p\left[\omega\right]\right)_1 
\left(r_\bk^p\left[\omega\right]\right)_2 e^{2ik_z L} 
\end{eqnarray}
$\rho_\bk^p$ and $f_\bk^p$ are respectively the open-loop and closed-loop functions
corresponding to one round trip in the cavity.
The system formed by the mirrors and fields is stable so that $f_\bk^p$ is an 
analytic function of frequency $\omega$. 
Analyticity is defined with the following physical conditions in the complex 
plane 
\begin{eqnarray}
\omega\equiv i\xi &&\quad,\quad \Re\xi > 0 \label{PhysConditions} \\
k_z\equiv i\kappa_\bk\left[\omega\right] \quad,\quad&& 
\kappa_\bk\left[\omega\right] \equiv\sqrt{\bk^2-\frac{\omega^2}{c^2}} 
\quad,\quad\Re\kappa_\bk\left[\omega\right] > 0 \nonumber 
\end{eqnarray}
The quantum numbers $p$ and $\bk$ remain spectator throughout the discussion 
of analyticity. The sum on transverse wavevectors may be represented as a sum 
over the eigenvectors $k_x=2\pi q_x/L_x,k_y=2\pi q_y/L_y$ associated with virtual 
quantization boxes along $x,y$ or, at the continuum limit $L_x,L_y\to\infty$
with $A=L_xL_y$, as an integral
\begin{eqnarray}
\sum_\bk &\equiv& \sum_{q_x=-\infty}^{\infty} \sum_{q_y=-\infty}^{\infty} 
\to A \int_{-\infty}^{\infty} \frac{\dd k_x}{2\pi}
\int_{-\infty}^{\infty} \frac{\dd k_y}{2\pi} 
\end{eqnarray}

We then introduce the Airy function defined in classical optics as the 
ratio of energy inside the cavity to energy outside the cavity for a given mode
\begin{eqnarray}
&&g_\bk^p\left[\omega\right] = 1 + \left\{ f_\bk^p\left[\omega\right] 
+ c.c. \right\} = \frac{1-\left|\rho_\bk^p\left[\omega\right]\right| ^2}
{\left| 1-\rho_\bk^p\left[\omega\right] \right| ^2} 
\label{Airyfunction}
\end{eqnarray}
As $f_\bk^p$, $g_\bk^p$ depends only on the reflection amplitudes of mirrors 
as they are seen from the inner side. 
We use a theorem which gives the commutators of the intracavity fields 
as the product of those well known for fields outside the cavity
by the Airy function.
This theorem was demonstrated with an increasing range of validity 
in \cite{Jaekel91}, \cite{Barnett98} and \cite{Genet03}. 
It is true regardless of whether the mirrors are lossy or not. 
Since it does not depend on the state of the field, it can
be used for thermal as well as vacuum fluctuations. 

Assuming thermal equilibrium, this theorem leads to the expression of 
the field anticommutators, \textit{i.e.} the field fluctuations.
Note that thermal equilibrium has to be assumed for the whole system,
which means that input fields as well as fluctuations associated with electrons, 
phonons and any loss mechanism inside the mirrors correspond 
to the same temperature $T$, whatever their microscopic origin may be. 
Then, the anticommutators of intracavity fields are given by those known 
for fields outside the cavity multiplied by the Airy function. 
Hence, the expression written in \cite{Genet03} for a null 
temperature is only modified through the appearance of a thermal factor 
in the integrand (compare with eq.(89) in \cite{Genet03} which is recovered
as the limit $\omega _\T \to 0$, that is also when the thermal factor 
$c\left[\omega\right]$ is replaced by unity)
\begin{eqnarray}
F &=&  -\hbar \sum_p \sum_\bk \int_{0}^{\infty} \frac{\dd\omega}{2\pi}
\left\{ i \kappa_\bk\left[\omega\right] f_\bk^p\left[\omega\right] 
c\left[\omega\right] + c.c. \right\} \nonumber \\
&&c\left[\omega\right] \equiv \coth\left( \frac{\pi\omega}{\omega_\T} \right) 
\quad,\quad \omega_\T \equiv \frac{2\pi k_\B T}{\hbar}  \label{ForceReal} 
\end{eqnarray}

Equation (\ref{ForceReal}) contains the contribution of ordinary modes freely
propagating outside and inside the cavity with $\omega > c \absk$ and $k_z$ real. 
This contribution thus merely reflects the intuitive picture of a radiation 
pressure of fluctuations on the mirrors of the cavity \cite{Jaekel91} with 
the factor $g_\bk^p-1$ representing a difference between inner and outer sides. 
Equation (\ref{ForceReal}) also includes the contribution of evanescent waves 
with $\omega < c \absk$ and $k_z$ imaginary. 
Those waves propagate inside the mirrors with an incidence angle larger than 
the limit angle and they also exert a radiation pressure on the mirrors,
due to the frustrated reflection phenomenon \cite{Genet03}.
Their properties are conveniently described through an analytical continuation 
of those of ordinary waves, using the well defined analytic behaviour of 
$\kappa_\bk$ and $f_\bk^p$. 

\section{The Casimir force as an integral over imaginary frequencies}

Using the same analyticity properties, we now transform
(\ref{ForceReal}) into an integral over imaginary frequencies
by applying the Cauchy theorem on the contour enclosing the quadrant 
$\Re\omega>0,\Im\omega>0$. We use high frequency transparency to neglect
the contribution of large frequencies. 

We deduce the following expression for the Casimir force 
\begin{eqnarray}
F = \hbar \sum_p\sum_\bk\int_{0}^{\infty}\frac{\dd\xi}{2\pi} 
\left\{ \kappa_\bk\left[i\xi+\eta\right] f_\bk^p \left[i\xi+\eta\right]  
c\left[i\xi+\eta\right] +c.c. \right\} &&
\label{ForceImag}
\end{eqnarray}
It is now written as an integral over complex frequencies $\Omega=i\xi +\eta$ 
close to the imaginary axis, with the small positive real number $\eta \to 0^+$ 
maintaining the Matsubara poles $\Omega_m=im\omega_\T$ of $c\left[\Omega\right]$
outside the contour used to apply the Cauchy theorem.
Up to this point, the present derivation is similar to Lifshitz' demonstration 
\cite{Lifshitz56} while being valid for arbitrary reflection amplitudes. 
The next steps in Lifshitz' derivation, scrutinized in the next section, 
may raise difficulties for arbitrary mirrors.

Before embarking in a closer inspection of this point, we give in 
the sequel of the present section another series expansion of (\ref{ForceImag}). 
It was already written in \cite{Genet00} for mirrors described
by the plasma model but is valid for arbitrary reflection amplitudes. 
It is based upon the expansion of the $\coth$ function into 
a series of exponentials
\begin{eqnarray}
&& c\left[i\xi+\eta\right] =1 + 
2\frac{\exp\left(-\frac{2\pi\left(i\xi+\eta\right)}{\omega_\T}\right)}
{1-\exp\left(-\frac{2\pi\left(i\xi+\eta\right)}{\omega_\T}\right)} 
= 2 \sum_n^\prime \exp\left(-\frac{2n\pi\left(i\xi+\eta\right)}{\omega_\T}\right)  
\label{cothExpand}
\end{eqnarray}
We have introduced the common summation convention 
\begin{eqnarray}
&&  \sum_n^\prime \varphi\left(n\right) \equiv \frac{1}{2}\varphi\left(0\right)
+\sum_{n=1}^{\infty}\varphi\left(n\right) \label{SumConv}
\end{eqnarray}

As a consequence of the presence of $\eta$,
the expansion (\ref{cothExpand}) is uniformly convergent so that,
when it is inserted in (\ref{ForceImag}), the order of the summation 
over $n$ and integration over $\xi$ may be exchanged.
It follows that the force (\ref{ForceImag}) may also be read as the following sum
\begin{eqnarray}
F&=& \frac\hbar\pi \sum_p\sum_\bk \sum_n^\prime 
\phit_\bk^p \left( \frac{2n\pi}{\omega_\T}\right) \label{ForceImagExp} \\
\phit_\bk^p \left(x\right) &\equiv& 2 \int_0^\infty \dd\xi 
\cos\left(\xi x\right) \phi_\bk^p \left[\xi\right]  \quad,\quad
\phi_\bk^p \left[\xi\right] \equiv \lim_{\eta\to 0^+} 
\kappa_\bk\left[i\xi+\eta\right] f_\bk^p \left[i\xi+\eta\right]
\nonumber 
\end{eqnarray}
The function $\phi_\bk^p$ is well defined almost everywhere, the only possible exception 
being the point $\xi=0$ where the limit $\eta \to 0^+$ may be ill defined
for mirrors described by dissipative optical models \cite{Klimchitskaya01}.
Since this is a domain of null measure,
the cosine Fourier transform $\phit_\bk^p$ of $\phi_\bk^p$ is well defined everywhere
and the expression (\ref{ForceImagExp}) of the Casimir force is valid for 
arbitrary mirrors, including dissipative ones.
Note that the term $n=0$ in (\ref{ForceImagExp}) corresponds exactly to the 
contribution of vacuum fluctuations, or to the zero temperature limit, while 
the terms $n\geq 1$ give the corrections associated with thermal fields.

\section{Comparison with Lifshitz' formula}

We come back to the derivation of the Lifshitz formula \cite{Lifshitz56}, often used as the 
standard expression of the Casimir force.
This formula is directly related to the decomposition of the $\coth$ function into elementary 
fractions corresponding to the Matsubara poles $\Omega_m=im\omega_\T$. 

Using this decomposition,
the factor $c\left[i\xi+\eta\right]$ appearing in (\ref{ForceImag}) is rewritten as 
a sum of Dirac delta functions and Cauchy principal values
\begin{equation}
c\left[i\xi+\eta\right] = \frac{\omega_\T}\pi \sum_{m=-\infty}^{+\infty}
\left\{  \pi\delta\left(\xi-m\omega_\T\right) -i \calP \frac{1}{\xi-m\omega_\T}\right\}
\label{DiracCauchy}
\end{equation}
If we assume furthermore that the function $\phi_\bk^p$ is a sufficiently smooth test function,
in the sense defined by the theory of distributions, we deduce that the expression
(\ref{ForceImag}) can also be read
\begin{equation}
F_\Lif = \frac{\hbar\omega_\T}\pi \sum_p\sum_\bk \sum_m^\prime 
\phi_\bk^p \left[m\omega_\T\right]
\label{LifshitzForm}
\end{equation}
This is the generalization of the Lifshitz' formula \cite{Lifshitz56} to the case of
arbitrary reflection amplitudes. It is a discrete sum over Matsubara 
poles with the primed summation symbol having the definition (\ref{SumConv}).
This formula is known to lead to the correct result in the case of dielectric mirrors
(for which it was derived in \cite{Lifshitz56}), for perfect mirrors \cite{Mehra67,Brown69} 
and also for metallic mirrors described by the lossless plasma model \cite{Genet00}.
However its applicability to arbitrary mirrors is a subject of controversy 
\cite{Klimchitskaya01}. 

In the context of the present paper, we want to discuss the mathematical conditions
of validity of the Lifshitz' formula with care.
For the derivation of (\ref{LifshitzForm}) to be valid, it is necessary that 
the function $\phi_\bk^p$ be a sufficiently smooth test function, so that 
it can be `multiplied' by expression (\ref{DiracCauchy}), in the sense
defined by the theory of distributions. 
Whether or not these smoothness conditions are obeyed at the point $\xi=0$ 
when $\phi_\bk^p$ is calculated from dissipative optical models constitutes the 
central question of the recent controversy on the value of the term $p$=TE, $m=0$ 
in Lifshitz' sum \cite{Bostrom00,Svetovoy00,Bordag00,Klimchitskaya01,Hoye03}. 
Let us repeat that, as stated in the preceding section, (\ref{ForceImagExp}) 
is still a mathematically valid expression of the Casimir force even when 
$\phi_\bk^p$ is ill defined for $\xi$ in a domain of null measure.

The question of validity of Lifshitz' formula (\ref{LifshitzForm}) may also be 
phrased in terms of applicability of the Poisson summation formula. 
As a matter of fact, the sum over $m$ in (\ref{LifshitzForm}) is a sum of regularly
spaced values of the function $\phi_\bk^p$ while the sum over $n$ in (\ref{ForceImagExp})
is a sum of regularly spaced values of its cosine Fourier transform $\phit_\bk^p$.
The identity between these two expressions is therefore ensured as soon as 
the Poisson summation formula \cite{MorseFeshbach} is true
\begin{eqnarray}
&&\sum_n^\prime \phit_\bk^p\left( \frac{2n\pi}{\omega_\T}\right) 
\overset{?}{=}
\omega_\T \sum_m^\prime \phi_\bk^p \left[m\omega_\T \right] \label{Poisson} 
\end{eqnarray}
It is known that this identity is true when $\phi_\bk^p$ is smooth enough. 
This condition is met for dielectric mirrors, for perfect mirrors and 
for mirrors described by the plasma model and this
explains why Lifshitz' formula (\ref{LifshitzForm}) may be used as 
well as (\ref{ForceImagExp}) in these cases \cite{Genet00}. 
Should this condition not be obeyed, Lifshitz' formula (\ref{LifshitzForm}) 
could give different results than the correct expression (\ref{ForceImagExp}).
For reflection amplitudes corresponding to a given optical model, the question 
whether or not (\ref{LifshitzForm}) is valid can only be decided through a detailed 
inspection of the mathematical conditions of validity of Poisson summation formula.

\section{Conclusion}

As explained in the Introduction, the aim of the present contribution was to 
re-derive the basic expression of the Casimir force between arbitrary mirrors 
at finite temperature. 
To this aim, we have repeated the derivation presented in \cite{Genet03} 
with the effect of thermal fluctuations now included. 
We have obtained two integral formulas (\ref{ForceReal},\ref{ForceImag}) of the 
force written over real and imaginary frequencies respectively. 
These formulas are identical to those obtained at zero temperature \cite{Genet03}, 
except for the appearance of the thermal coth factor.
They are also identical to those already published in \cite{Jaekel91} for
the restricted case of non dissipative mirrors. 
As in \cite{Jaekel91}, the mirrors have been described by frequency dependent 
reflection amplitudes obeying general properties of scattering theory but 
lossy mirrors are now considered on the same footing as lossless ones,
with additional fluctuations accompanying losses properly accounted for.

The two integral formulas (\ref{ForceReal},\ref{ForceImag}) are mathematically 
equivalent and they represent the prediction of Quantum ElectroDynamics (QED) 
for the Casimir force between arbitrary mirrors at non zero temperature, 
as soon as the original Casimir geometry is considered. 
Comparison of this QED prediction with experimental results would be a direct test
of the quantum theory of mechanical effects of vacuum and thermal field fluctuations 
if the reflection amplitudes of the mirrors were measured simultaneously.
In fact, the integral (\ref{ForceImag}) is usually calculated
for mirrors described by optical models. The reflection amplitudes are calculated from 
the Fresnel laws for a bulk mirror 
\begin{eqnarray}
r_\bk^\TE\left[\omega\right] &=& \frac{\kappa_\bk\left[\omega\right]-\kappam_\bk\left[\omega\right]}%
{\kappa_\bk\left[\omega\right]+\kappam_\bk\left[\omega\right]} \quad,\quad
r_\bk^\TM\left[\omega\right] = \frac{\kappam_\bk\left[\omega\right]%
-\varepsilon\left[\omega\right]\kappa_\bk\left[\omega\right]}%
{\kappam_\bk\left[\omega\right]+\varepsilon\left[\omega\right]%
\kappa_\bk\left[\omega\right]} \nonumber \\
&&\kappam_\bk\left[\omega\right] =\sqrt{\bk^2-\varepsilon\left[\omega\right]\frac{\omega^2}{c^2}} 
\quad,\quad \Re\kappam_\bk\left[\omega\right] >0 \label{FresnelLaws} 
\end{eqnarray}
$\kappa_\bk$ is representing $k_z$ in vacuum (see eq.(\ref{PhysConditions})) 
and $\kappam_\bk$ is analogously defined in the bulk.
The dielectric function $\varepsilon$ is often chosen to fit the Drude model 
\begin{eqnarray}
\varepsilon\left[\omega\right]=1-
\frac{\omega_\pl^2}{\omega\left(\omega+i\gamma_\D\right)}
\label{Drude} 
\end{eqnarray}
The lossless plasma model corresponds to $\gamma_\D \to 0$ in (\ref{Drude}).
Clearly, a non null value of $\gamma_\D$ is a better description of real metals than a 
null one, as shown in particular by the study of tabulated optical data \cite{Lambrecht00}.
In this context, \textit{ad hoc} prescriptions sometimes used to bypass the difficulties 
associated with dissipation are of little help \cite{Klimchitskaya01}. 
What is needed is the theoretical expression of the force between real mirrors, 
the optical response of which certainly involves dissipation of conduction electrons. 

In order to contribute to the resolution of this difficulty, we have discussed two 
expressions derived from (\ref{ForceImag}) through expansions of the thermal coth factor. 
The first expression (\ref{ForceImagExp}) relies on a uniformly convergent expansion
of this factor in terms of exponentials. The second one is the Lifshitz' formula 
(\ref{LifshitzForm}) obtained from an expansion of the same factor over its poles.
The first expression may have a wider scope of validity than the second one with
the identity being ensured as soon as the function $\phi_\bk^p$ verifies the
Poisson summation formula.
At the end of this contribution, we are thus confronted with the following alternative.
Either Poisson summation formula is valid and so is Lifshitz' expression,
or Poisson formula is not valid because the function $\phi_\bk^p$ is not smooth enough 
and, therefore, Lifshitz' formula may fail to be a correct expression of the Casimir force.
In both cases however, the integral expressions (\ref{ForceImag},\ref{ForceImagExp})
of the Casimir force, which have been demonstrated by using scattering theory, remain valid.

\section*{Acknowledgements}

We thank G. Barton, E. Fischbach and U. Mohideen for stimulating discussions.

\newcommand{\Review}[1]{\textrm{#1}}
\newcommand{\Volume}[1]{\textbf{#1}}
\newcommand{\Book}[1]{\textit{#1}}
\newcommand{\Eprint}[1]{\textsf{#1}}

\end{document}